# Efficient Optimal Algorithm of Task Scheduling in Cloud Computing Environment

Dr. Amit Agarwal, Saloni Jain

*(Department of Computer Science University of Petroleum and Energy, Dehradun, India)*
*(M.Tech in Computer Science and Engineering Sharda University, Greater Noida, India)*

***Abstract:*** *Cloud computing is an emerging technology in distributed computing which facilitates pay per model as per user demand and requirement. Cloud consist of a collection of virtual machine which includes both computational and storage facility. The primary aim of cloud computing is to provide efficient access to remote and geographically distributed resources.*

*Cloud is developing day by day and faces many challenges, one of them is scheduling. Scheduling refers to a set of policies to control the order of work to be performed by a computer system. A good scheduler adapts its scheduling strategy according to the changing environment and the type of task. In this research paper we presented a Generalized Priority algorithm for efficient execution of task and comparison with FCFS and Round Robin Scheduling. Algorithm should be tested in cloud Sim toolkit and result shows that it gives better performance compared to other traditional scheduling algorithm.*

***Index Terms****- Virtual Machine, Scheduling, Cloud Computing*

## 1. INTRODUCTION

Cloud computing comes in focus development of grid computing, virtualization and web technologies. Cloud computing is an internet based computing that delivers infrastructure as a service (IaaS), platform as a service (PaaS), and software as services (SaaS). In SaaS, software application is made available by the cloud provider. In PaaS an application development platform is provided as a service to the developer to create a web based application. In IaaS computing infrastructure is provided as a service to the requester in the form of Virtual Machine (VM).These services are made available on a subscription basis using pay-as-you-use model to customers, regardless of their location. Cloud Computing still under in its development stage and has many issues and challenges out of the various issues in cloud scheduling plays very important role in determining the effective execution.

Scheduling refers to the set of policies to control the order of work to be performed by a computer system. There has been various types of scheduling algorithm existing in distributed computing system, and job scheduling is one of them. The main advantage of job scheduling algorithm is to achieve a high performance computing and the best system throughput. Scheduling manages availability of CPU memory and good scheduling policy gives maximum utilization of resource. We compared three algorithm Time Shared, Space shred and generalizes priority algorithm.

## 2. RELATED WORK

In this section, we describe the related work ok task scheduling in cloud computing environment. The author of paper [1] presented a brief description of cloud Sim toolkit and his Functionality. Cloud Sim toolkit is a platform where you can test your work before applied into real work, in this paper we learned how to simulate a task with different approaches and different scheduling policy.

In paper [2] author proposed an approach for task scheduling algorithm based on load balancing in cloud computing. This paper described two level task scheduling based on the load balancing. This type of task scheduling cannot only meet user's requirement but also provide high resource utilization. This paper presented the implementation of an efficient Quality of Service (QoS) based Meta-Scheduler and Backfill strategy based light weight Virtual Machine Scheduler for dispatching jobs

The authors of paper [3] presented an optimized algorithm for task scheduling based on genetic simulated annealing algorithm. This considers the QoS requirements like completion time, bandwidth, cost, distance, reliability of different type tasks. Here annealing is implemented after the selection, crossover and mutation, to improve local search ability of genetic algorithm.

In this paper [4] hierarchical scheduling is presented which helps in achieving Service Level Agreement with quick response from the service provider. In our proposed approach Quality of Service metric such as response time is achieved by executing the high priority jobs (deadline based jobs) first by estimating job completion time and the priority jobs are spawned from the remaining job with the help of Task Scheduler.

In paper [5] author presented an optimized algorithm for task scheduling based on Activity Based Costing (ABC). This algorithm assigns priority level for each task and uses cost drivers. ABC measures both cost of the object and performance of the activities.

The paper [6] presented transaction intensive cost constraint cloud Work flow scheduling algorithm. Algorithm consider execution cost and execution time as the two key considerations. The algorithm minimize the cost under certain user designated deadlines. Our proposed methodology is mainly based on computational capability of Virtual Machines.

In paper [7] a new VM Load Balancing Algorithm is Weighted Active Monitoring Load Balancing Algorithm using CloudSim tools, for the Datacenter to effectively load balance requests between the available virtual machines assigning a weight, in order to achieve better performance parameters. Here VMs of different processing powers and the tasks/requests are assigned or allocated to the most powerful VM and then to the lowest and so on

In paper [8] author proposed an algorithm is Ant colony optimization in which random optimization search approach is used for allocating the incoming jobs to the virtual machines This algorithm uses a positive feedback mechanism and imitates the behavior of real ant colonies in nature to search for food and to connect to each other by pheromone laid on paths traveled.

In paper [9] is analyzing and evaluating the performance of various CPU scheduling in cloud environment using Cloud Sim the basic algorithm OS like FCFS, Priority Scheduling and Shortest Job First , we test under different which scheduling policy perform better .

In [10] author proposes a priority based dynamic resource allocation in cloud computing. This paper considers the multiple SLA parameter and resource allocation by pre-emption mechanism for high priority task execution can improve the resource utilization in cloud. The main highlight of the paper is that it provides dynamic resource provisioning and attains multiple SLA objectives through priority based scheduling. Since cost is the important aspect in cloud computing

## 3. PROPOSED FRAME WORK AND METHODOLOGY

Resource allocation and scheduling of resources have been an important aspect that affects the performance of networking, parallel, distributed computing and cloud computing. Many researchers have proposed various algorithms for allocating, scheduling and scaling the resources efficiently in the cloud. Scheduling process in cloud can be generalized into three stages namely–

**Resource discovering and filtering** – Datacenter Broker discovers the resources present in the network system and collects status information related to them.

**Resource selection** – Target resource is selected based on certain parameters of task and resource. This is deciding stage.

**Task submission** -Task is submitted to resource selected.

Here we mainly discuss three scheduling algorithm First come first serve, Round robin scheduling and new scheduling approach is generalized priority algorithm.

### 3.1 First come first serve-

FCFS for parallel processing and is aiming at the resource with the smallest waiting queue time and is selected for the incoming task. The Cloud Sim toolkit supports First Come First Serve (FCFS) scheduling strategy for internal scheduling of jobs. Allocation of application-specific VMs to Hosts in a Cloud-based data center is the responsibility of the virtual machine provisioned component. The default policy implemented by the VM provisioned is a straightforward policy that allocates a VM to the Host in First-Come-First-Serve (FCFS) basis. The disadvantages of FCFS is that it is non preemptive. The shortest tasks which are at the back of the queue have to wait for the long task at the front to finish .Its turnaround and response is quite low.

### 3.2 Round Robin Scheduling-

Round Robin (RR) algorithm focuses on the fairness. RR uses the ring as its queue to store jobs. Each job in a queue has the same execution time and it will be executed in turn. If a job can't be completed during its turn, it will be stored back to the queue waiting for the next turn. The advantage of RR algorithm is that each job will be executed in turn and they don't have to be waited for the previous one to get completed. But if the load is found to be heavy, RR will take a long time to complete all the jobs. The Cloud Sim toolkit supports RR scheduling strategy for internal scheduling of jobs. The drawback of RR is that the largest job takes enough time for completion.

### 3.3 Generalized Priority Algorithm-

Customer define the priority according to the user demand you have to define the parameter of cloudlet like size, memory, bandwidth scheduling policy etc. In the proposed strategy, the tasks are initially prioritized according to their size such that one having highest size has highest rank. The Virtual Machines are also ranked (prioritized) according to their MIPS value such that the one having highest MIPS has the highest rank. Thus, the key factor for prioritizing tasks is their size and for VM is their MIPS. This policy is performing better than FCFS and Round Robin scheduling.

Consider a 5 computational specific Virtual Machines represented by their Id and MIPS
As V = {{0, 250}, {1, 1000}, {2, 250}, {3, 500}, {4, 250}}. Here Vm2 will get first preference because of the highest MIPS, second preference is given to Vm4 and then Vm1, Vm3 and Vm4 get rest preferences.

**Algorithm-**

The algorithm is given below. This algorithm stores all suitable Virtual Machines in a VM List.
─────────────────────────────────
prev −99
push first vertex
while Stack 6= Empty do
get unvisited vertex adjacent to stack top
if no adjacent vertex then
if prev 6= StackTop then
copy all stack contents to VM List
end if
pop
if Stack 6= Empty then
prev = StackTop
end if
else
mark the node as visited
push adjacent vertex
end if
end while

Here are

**Step –1** Create VM to different Datacenter according to computational power of host/physical server in term of its cost processor, processing speed, memory and storage.

**Step-2** Allocate cloudlet length according to computational power.

**Step -3** Vm Load Balancer maintain an index table of Vms, presently vm has zero allocation.

**Step -4** Cloudlet bound according to the length and respective MIPS.

**Step -5** Highest length of cloudlet get highest MIPS of virtual machine.

**Step -6** Datacenter broker sends the request to the Vm identified with id

**Step -7** Update the available resource.

## 4. EXPERIMENT AND EVALUATION

In order to verify our algorithm, we conducted experiment on Intel(R) core(TM) i5 Processor 2.6 GHz, Windows 7 platform and using CloudSim 3.0.3 simulator. The Cloud Sim toolkit supports modeling of cloud system components such as data centers, host, virtual machines, scheduling and resource provisioning policies. A tool kit is the utilization which open the possibility of evaluating the hypothesis prior to software development in an environment where one can reproduce tests We have created 5 Virtual Machines using Vm component and set the property of RAM as 512 MB for all Virtual Machines, and the MIPS as 250, 1000, 250, 500 and 250 respectively. We have created 12 tasks using Cloudlet component and set the property of Cloudlet length as 20000, 10000, 20000, 10000, 10000, 20000, 10000, 20000, 10000,10000, 20000 and 10000 respectively. For this we considered 5 Virtual Machines with MIPS 1000, 500, 250, 250, 250 and RAM size of all Virtual Machine as 512 MB. Experiment is conducted for varying number of tasks like 100, 200, 300, 400 and 500 respectively. For comparison and analysis, we implemented the FCFS , Round robin, Generalized priority algorit

| FCFS Algorithm | | | Round Robin Algorithm | | | Generalized Priority Algorithm | | |
|---|---|---|---|---|---|---|---|---|
| Execution Time | Data Center id | VM | Execution Time | Data Center id | VM | Execution Time | Data Center id | VM |
| 80 | 2 | 1 | 239.99 | 2 | 1 | 20 | 2 | 2 |
| 10 | 2 | 2 | 119.99 | 2 | 2 | 20 | 3 | 1 |
| 80 | 2 | 3 | 160 | 2 | 3 | 40 | 3 | 4 |
| 20 | 3 | 4 | 40 | 2 | 4 | 40 | 2 | 3 |
| 40 | 3 | 5 | 20 | 3 | 5 | 40 | 3 | 5 |
| 80 | 2 | 1 | 239.99 | 2 | 1 | 20 | 2 | 2 |
| 10 | 2 | 2 | 119.99 | 2 | 2 | 40 | 2 | 1 |
| 80 | 2 | 3 | 160 | 2 | 3 | 40 | 3 | 4 |
| 20 | 3 | 4 | 40 | 2 | 4 | 40 | 2 | 3 |
| 40 | 3 | 5 | 20 | 3 | 5 | 40 | 3 | 5 |
| 80 | 2 | 1 | 239.99 | 2 | 1 | 20 | 2 | 2 |
| 10 | 2 | 2 | 119.99 | 2 | 2 | 10 | 2 | 4 |
| 45.8 | | | 114.16 | | | 30 | | |

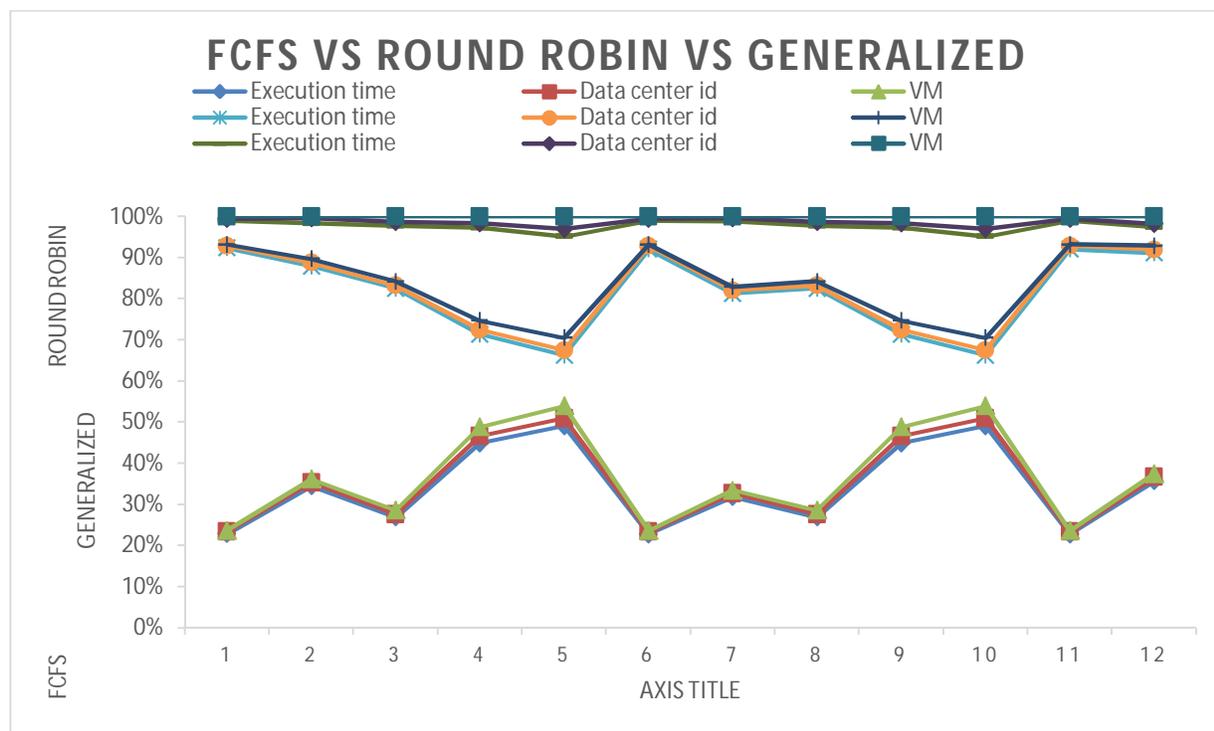

## 5. CONCLUSION

Scheduling is one of the most important tasks in cloud computing environment. In this paper we have analyzed various scheduling algorithm which efficiently schedules the computational tasks in cloud environment. We have created FCFS, Round robin scheduling Algorithm and new proposed Scheduling algorithm is (GPA) generalized priority algorithm. Priority is an important issue of job scheduling in cloud environments. The experiment is conducted for varying number of Virtual Machines and workload traces. The experiment conducted is compared with FCFS and Round Robin. The result shows that the proposed algorithm is more efficient than FCFS and Round Robin algorithm.

## 6. FUTURE WORK

In This paper we mainly discuss three algorithm we developed a new generalized priority based algorithm with limited task, future we will take more task and try to reduce the execution time as presented and we develop this algorithm to grid environment and will observe the difference of time in cloud an grid.

## REFERENCES

[1] Burya R Raman, R. Calheiros, R.N.(2009) "*Modeling and Simulation of Scalable Cloud Environment and the Cloud Sim Toolkit: Challenges and Opportunities*'', IEEE publication 2009,pp1-11

[2] Dr. Sudha Sadhasivam, R. Jayarani, Dr. N. Nagaveni, R. Vasanth Ram "*Design and Implementation of an efficient Two-level Scheduler for Cloud Computing Environment*" In

Proceedings of International Conference on Advances in Recent Technologies in Communication and Computing, 2009

[3] G. Guo-Ning and H. Ting-Lei, "*Genetic Simulated Annealing Algorithm for Task Scheduling based on Cloud Computing Environment,*" In Proceedings of International Conference on Intelligent Computing and Integrated Systems, 2010, pp. 60-63

[4] Rajkumar Rajavel , Mala T "*Achieving Service Level Agreement in Cloud Environment using Job Prioritization in Hierarchical Scheduling*" Proceeding of International Conference on Information System Design and Intelligent Application,2012 , *vol 132*, pp 547-554

[5] Q. Cao, W. Gong and Z. Wei, "*An Optimized Algorithm for Task Scheduling Based On Activity Based Costing in Cloud Computing,*" In Proceedings of Third International Conference on Bioinformatics and Biomedical Engineering, 2009*, pp. 1-3*

[6] Y. Yang, Kelvin, J. chen, X. Lin, D.Yuan and H. Jin, "*An Algorithm in Swin DeW-C for Scheduling Transaction Intensive Cost Constrained Cloud Workflow,*" In Proceedings of Fourth IEEE International Conference on eScience, 2008, *pp. 374-375*

[7] Jasmin James, Dr. Bhupendra Verma "*Efficient Vm Load Balancin Algorithim For A Cloud Computing Environment* " In Proceeding of International Journal on Computer Science and Engineering (IJCSE) *Vol. 4 No. 09,*  Sep 2012

[8] Medhat A. Tawfeek, Ashraf El-Sisi, Arabi E. keshk, Fawzy A. Torkey "*Cloud Task Scheduling Based on Ant Colony Optimization*" In Proceeding of IEEE International Conference on Computer Engineering & Systems (ICCES), 2013

[9] Monica Gahlawat, Priyanka Sharma (2013) *" Analysis and Performance Assessment of CPU Scheduling Algorithm in  Cloud Sim*" International Journal of Applied Information System(IJAIS)-*1SSN: 2249-0868* Foundation of Computer Science FCS, New York, USA *Volume5- No 9*, July 2013

[10]  Pawar, C. S., & Wagh, R. B. (2012). *"Priority Based Dynamic resource allocation in Cloud computing"*, International Symposium on Cloud and Services Computing, IEEE, 2012 *pp 1-6*

[11]  Raghavendra  Achar_, P. Santhi Thilagam, Shwetha D_, Pooja H_, Roshni_ and Andrea "*Optimal Scheduling of Computational Task inCloud using Virtual Machine Tree*" In Proceeding of Third International Conference on Emerging Applications of Information Technology (EAIT), IEEE Publication , 2012

[12] Gemma Reig, Javier Alonso and Jordi Guitart, "*Prediction of Job Resource Requirements for Deadline Schedulers to Manage High-Level SLAs on the Cloud*", 9th IEEE International Symposium on Network Computing and Applications, 2010

[13] Jinhua Hu, Jianhua Gu, Guofei Sun, Tianhai Zhao, NPU HPC Center Xi'an, China "*A Scheduling Strategy on Load Balancing of Virtual Machine Resources in Cloud Computing Environment*", IEEE 2010

[14] Suraj Pandey, Department of Computer Science and Software Engineering, the University of Melbourne, Australia, "*Scheduling and Management of Data Intensive Application Workflows in Grid and Cloud Computing Environments*", Dec 2010

[15] Ashutosh Ingole, Sumit Chavan, Utkarsh Pawde. "*An optimized algorithm for task scheduling based on activity based costing in cloud computing"* (NCICT) 2011, Proceedings published in International Journal of Computer Applications® (IJCA)

[16] Zhong, H., Tao, K. and Zhang, X. 2010 " An Approach to Optimize Resource Scheduling Algorithm for Open-Source Cloud Systems"  The Fifth Annual China Grid Conference. IEEE Computer  Society, 978-0-7695-4106-8.